\documentclass{article}
\usepackage{spconf,amsmath,graphicx,hyperref}
\usepackage{multirow}
\usepackage{tabularx}
\usepackage{enumitem}
\usepackage{booktabs}
\usepackage[dvipsnames]{xcolor}
\usepackage{pifont}

\newcommand{\cmark}{{\begingroup\color{ForestGreen}\ding{51}\endgroup}}
\newcommand{\xmark}{{\begingroup\color{BrickRed}\ding{55}\endgroup}}

\newcommand{\ymark}{{\begingroup\ding{51}\endgroup}}
\newcommand{\wmark}{{\begingroup\ding{55}\endgroup}}

\usepackage{tikz}
\newcommand{\halfcheck}{%
  \tikz[baseline=-0.5ex]{%
    \node[inner sep=0pt] (a) {\textcolor{ForestGreen}{\ding{51}}};
    \draw[BrickRed, thick] (a.north west) -- (a.south east);
  }%
}

\title{ISSE: An Instruction-Guided Speech Style Editing Dataset and Benchmark}
\name{Yun Chen\textsuperscript{1}, Qi Chen\textsuperscript{2}, 
Zheqi Dai \textsuperscript{3}, 
Arshdeep Singh\textsuperscript{1}, Philip J.B. Jackson\textsuperscript{1}, Mark D. Plumbley\textsuperscript{1,*}\thanks{* Corresponding author.}}
\address{\textsuperscript{1}Centre for Vision Speech and Signal Processing, University of Surrey, United Kingdom \\ \textsuperscript{2}ByteDance Intelligent Creation, China, \ \ \ \ \textsuperscript{3}The Chinese University of Hong Kong, China}

\begin{document}
\ninept
\maketitle
\begin{abstract}
Speech style editing refers to modifying the stylistic properties of speech while preserving its linguistic content and speaker identity. 
However, most existing approaches depend on explicit labels or reference audio, which limits both flexibility and scalability. 
More recent attempts to use natural language descriptions remain constrained by oversimplified instructions and coarse style control.
To address these limitations, we introduce an \textbf{I}nstruction-guided \textbf{S}peech \textbf{S}tyle \textbf{E}diting Dataset (ISSE). 
The dataset comprises nearly 400 hours of speech and over 100,000 source-target pairs, each aligned with diverse and detailed textual editing instructions.
We also build a systematic instructed speech data generation pipeline leveraging large language model, expressive text-to-speech and voice conversion technologies to construct high-quality paired samples. 
Furthermore, we train an instruction-guided autoregressive speech model on ISSE and evaluate it in terms of instruction adherence, timbre preservation, and content consistency. 
Experimental results demonstrate that ISSE enables accurate, controllable, and generalizable speech style editing compared to other datasets.
The project page of ISSE is available at \href{https://ychenn1.github.io/ISSE/}{https://ychenn1.github.io/ISSE/}.
\end{abstract}
\begin{keywords}
ISSE dataset, Speech style editing, Instruction-guided editing, Natural language instructions
\end{keywords}

\section{Introduction}
\label{sec:intro}

Speech style editing refers to the task of modifying the stylistic properties of speech, such as emotional tone, speaking rate, and prosodic delivery, while preserving its linguistic content and speaker identity ~\cite{zuo2025enhancing,huang2024instructspeech}.
Existing approaches in this domain largely rely on explicit labels~\cite{zhou2020converting} or reference audio~\cite{chen23_interspeech, zhang2025vevo} to provide editing signals, which constrains their flexibility and scalability~\cite{guo2023prompttts}. 
With the rapid advances in natural language modeling \cite{touvron2023llama, bai2023qwen}, a new paradigm has emerged: guiding speech generation through natural language descriptions \cite{guo2023prompttts,niu2024hybridvc, yao2024promptvc}. 

\begin{figure}[t]
    \centering
    \includegraphics[width=1\linewidth]{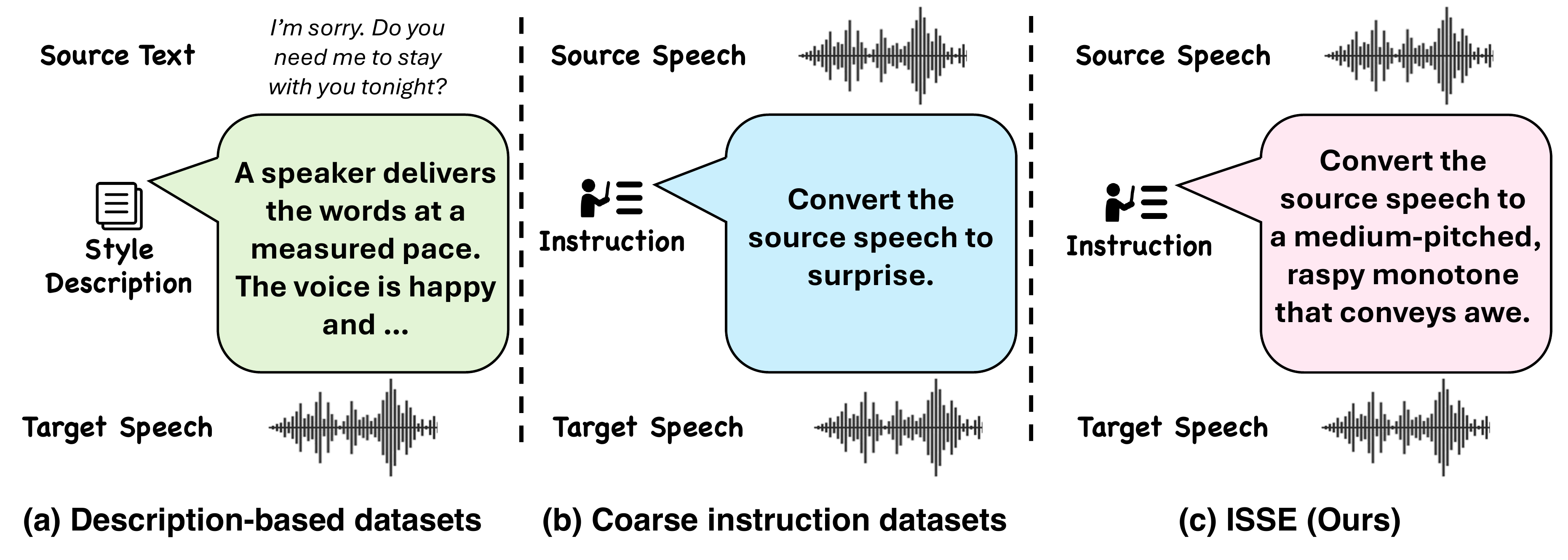}
    \vspace{-10pt}
    \caption{\textbf{Comparison of ISSE with existing speech generation datasets.} (a) Description-based expressive speech datasets provide only text and style descriptions. (b) Coarse instruction datasets provide paired source-target speech with simple template instructions, but remain limited for fine control. (c) Our proposed ISSE provides paired source-target speech with fine-grained style instructions.}
    \label{fig:comparation}
    \vspace{-10pt}
\end{figure}

\begin{table*}[t]
    \centering
     \setlength{\tabcolsep}{7pt}
    \caption[]{\textbf{The comparative analysis of our dataset against previous speech style datasets.} \halfcheck \ indicates that only part of the speeches meet the condition and * denotes style version of the dataset.}
    \begin{tabular}{l|ccccc|c}
       \toprule
       Dataset & Paired speech & \# Styles & Description type & Instruction & Duration & Open-source \\
       \midrule
        PromptSpeech \cite{guo2023prompttts}  & \xmark &  - & Natural language & None & Unavailable &  \halfcheck \\
        CapSpeech* \cite{wang2025capspeech} & \xmark & 28 & Natural language & None & 90h & \cmark\\ 
        Ears \cite{richter2024ears} & \xmark & 22 & Label & None & 100h & \cmark  \\
        Expresso \cite{nguyen2023expresso} & \halfcheck & 9 & Label & None & 42h total / 11h paired & \cmark \\
        ESD \cite{zhou2022emotional} & \cmark  & 5  & Label & None & 29h & \cmark\\
        InstructSpeech \cite{huang2024instructspeech} & \cmark & 22 & Natural language & Coarse-grained & Unavailable & \xmark\\
        \midrule
        ISSE (Ours) & \cmark & 28 & Natural language & Fine-grained & 382h  & \cmark\\
        \bottomrule
    \end{tabular}
    \label{tab:dataset_comparation}
    \vspace{-2pt}
\end{table*}

A key prerequisite for training text-guided speech generation models is the availability of paired “speech-text description” datasets. 
In the context of text-to-speech (TTS), several studies have annotated speech with textual style descriptions to produce more expressive outputs \cite{guo2023prompttts,wang2025capspeech,jin2024speechcraft}, as shown in~Fig.~\ref{fig:comparation} (a). 
However, these datasets are constructed for stylized speech generation, which cannot describe the change between source speech and target speech when applying it to the style editing task.
This makes the editing objective ill-posed and hindering the disentanglement of style from content and timbre~\cite{niu2024hybridvc,yao2024promptvc}.
To enable attribute-specific editing, Huang et al. \cite{huang2024instructspeech} constructed a dataset of source-target speech pairs with corresponding editing instructions, as illustrated in Fig.~\ref{fig:comparation}~(b). 
Despite its contribution, these datasets exhibit two notable limitations. 
\textbf{First, existing datasets suffer from limitations in the design of instructions.}
The provided instructions are often overly simplistic and lack linguistic diversity, typically restricted to templated commands such as “Change the emotion of speech” or “Change the speaking speed of speech.” 
Moreover, they are usually coarse-grained, focusing on only a single attribute to modify.
Such constraints hinder the ability to represent more realistic and expressive editing scenarios.
In contrast, fine-grained instructions (e.g., “Convert the source speech to a medium-pitched, raspy monotone that conveys awe”) help enable flexible, multi-attribute speech style editing, as shown in Fig.~\ref{fig:comparation}~(c).
\textbf{Secondly, most existing open-source paired speech style datasets remain limited in scale and diversity. }
As shown in~Table~\ref{tab:dataset_comparation}, ESD~\cite{zhou2022emotional} contains only 29 hours of recordings, while other datasets such as Ears~\cite{richter2024ears} and Expresso~\cite{nguyen2023expresso} are similarly constrained. 
These datasets not only suffer from relatively small amounts of data, but also lack sufficient variety in speaking styles, prosodic patterns, and content coverage. 
Such limitations in both quantity and diversity pose significant challenges and have become a major bottleneck for advancing research in speech style editing.

\begin{figure*}[!t]  
    \centering
    \includegraphics[width=0.95\linewidth]{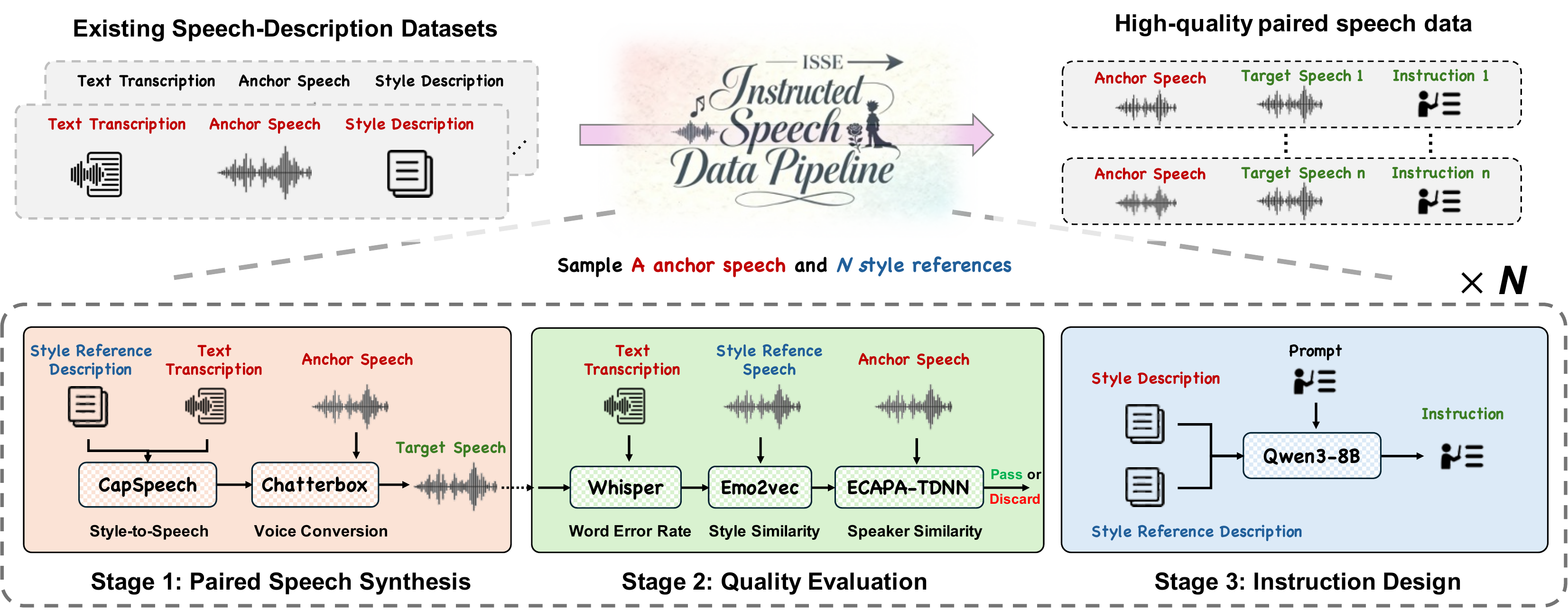}
    \vspace{-3pt}
    \caption{
    \textbf{Framework of the proposed instructed speech data pipeline.} 
It contains three stages: (1) \textit{Paired Speech Synthesis}, which generates \textcolor{ForestGreen}{target speeches} by combining the \textcolor{BrickRed}{anchor speech} (providing linguistic content and timbre) with styles from \textcolor{MidnightBlue}{style references}; 
(2) \textit{Quality Evaluation}, where \textcolor{ForestGreen}{target speeches} are assessed in terms of word error rate, style similarity, and speaker similarity, and only those that meet the criteria are retained;
and (3) \textit{Instruction Design}, which employs an LLM to produce editing instructions that link the \textcolor{BrickRed}{anchor speech} and the \textcolor{ForestGreen}{target speech}. 
For each \textcolor{BrickRed}{anchor speech}, $N$ \textcolor{MidnightBlue}{style references} are sampled, yielding up to $N$ paired speeches in a single sampling process.
}
    \label{fig:fig1}
    \vspace{-8pt}
\end{figure*}

To address these challenges, we present ISSE, a novel \textbf{I}nstruction-guided \textbf{S}peech \textbf{S}tyle \textbf{E}diting dataset.
ISSE contains nearly 400 hours of speech and over 100,000 source-target pairs, and is designed with fine-grained editing instructions, substantially exceeding the simplicity and limited coverage of prior resources (Table~\ref{tab:dataset_comparation}).
As the performance of style editing models hinges on the consistency between fine-grained instructions and the paired speech content, we design an effective data construction pipeline.
Specifically, we integrate expressive TTS and voice conversion (VC) to generate reliable source-target pairs that preserve content and timbre while varying style attributes.
We further conduct quality evaluation of content preservation, speaker identity consistency, and style similarity to filter low-quality data.
Finally, a large language model (LLM) is employed to describe the transformation between paired speech samples and generate the fine-grained instruction.
To validate the effectiveness of the proposed ISSE, we establish a benchmark using a baseline instruction-guided autoregressive speech model, LlasaEdit. 
We evaluate LlasaEdit trained on ISSE against the same model trained on ESD under both in-domain and cross-domain settings. 
The experimental results consistently confirm the effectiveness of ISSE for instruction-guided speech style editing.

\begin{itemize}[leftmargin=1em,itemsep=0.1em,topsep=0.2em]
\item We release the ISSE dataset, which contains approximately 400 hours of speech and more than 100,000 source-target pairs annotated with fine-grained editing instructions.
\item We propose an instructed speech data pipeline that integrates existing speech datasets with TTS, VC, quality assessment, and LLM tools to generate high-quality speech pairs with fine-grained editing instructions.
\item We establish a benchmark with an instruction-guided autoregressive model trained on ISSE. Experiments demonstrate the effectiveness of both the dataset and the proposed model in editing style while preserving transcript and identity.
\end{itemize}

\begin{figure*}[!t]
    \centering
\includegraphics[width=1\linewidth]{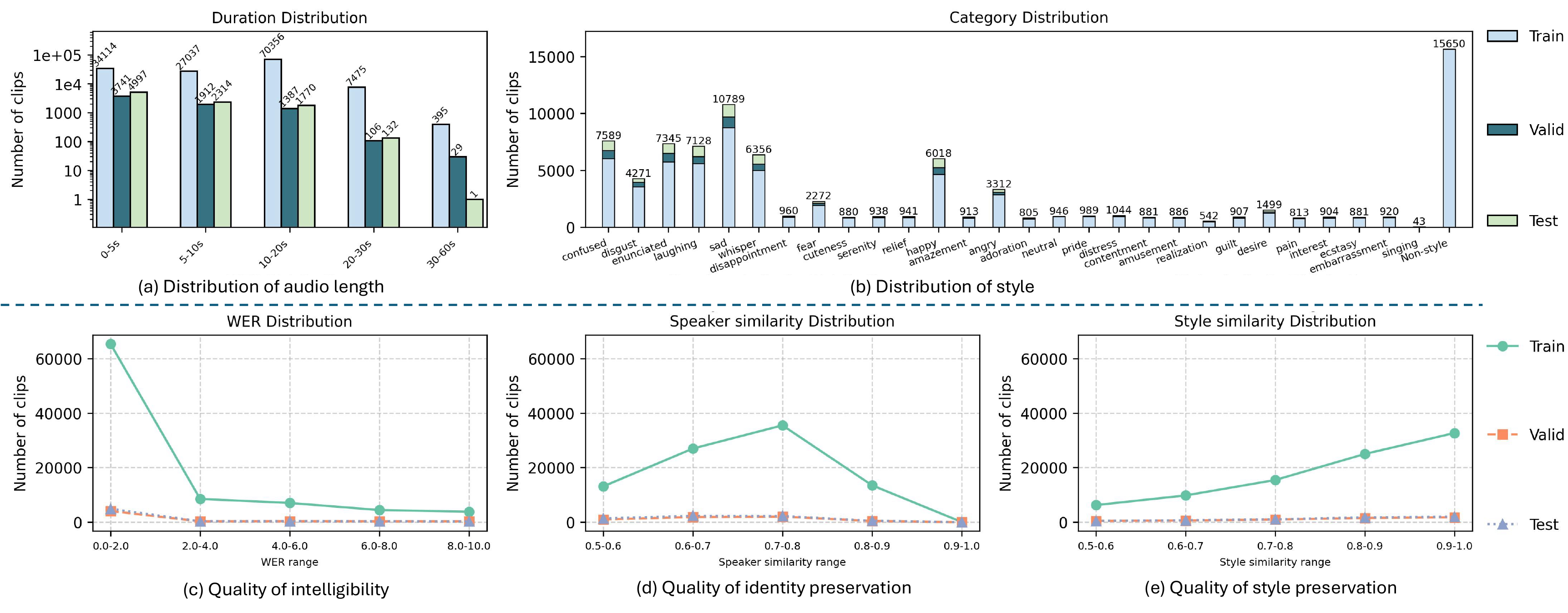}
    \vspace{-10pt}
    \caption{\textbf{Data distribution and quality evaluation in the ISSE dataset.} (a) Audio length distribution and (b) style distribution demonstrate balanced coverage and diverse stylistic attributes. (c)-(e) show quality assessments of generated samples, including intelligibility, identity preservation, and style preservation, confirming the reliability of the generated speeches.}
    \vspace{-10pt}
    \label{fig:style_Distributions}
\end{figure*}

\section{Dataset Construction}
\label{sec:format}
In this section, we propose an instructed speech data pipeline to generate source and target speech pairs with corresponding style instructions.
As shown in~Fig.~\ref{fig:fig1}, the pipeline consists of three stages of data processing: 
\textbf{Stage 1: Paired Speech Generation}: Generates speech samples in diverse styles and then converts them to share the same speaker identity while maintaining the style differences; 
\textbf{Stage 2: Quality Evaluation}: Validates the generated speech quality and consistency based on predefined criteria;
\textbf{Stage 3: Instruction Design}: Formulates natural language instructions describing the style transformation between the paired speech samples.

\subsection{Paired Speech Generation}
We adopt two existing speech datasets, \textit{Ears} \cite{richter2024ears} and \textit{Expresso} \cite{nguyen2023expresso}, as the source data.
Both datasets contain speeches, transcriptions, and the corresponding style descriptions annotated with the tool used in \cite{lyth2024natural}.
To minimize the influence of speaker identity during data construction, non-stylistic words (e.g., he, she) are removed from the style descriptions.
From the processed datasets, we first sample an anchor speech, with its transcription and speaker identity serving as fixed attributes for the subsequent synthesis process.
We then randomly select $n$ additional samples with distinct speaking styles, referred to as style reference speeches.
Their style descriptions are used to specify the target styles for synthesis.
To produce expressive speech variations, we employ EmoCapTTS \cite{wang2025capspeech}.
For each case, EmoCapTTS takes as input the transcription of the anchor speech and the style description of a style reference speech, and generates a stylized target that preserves the linguistic content while adopting the style of the style reference speech. 

The EmoCapTTS model lacks explicit control over speaker identity, so the generated speech differs from the anchor speech in timbre.
This variability makes it difficult to design editing instructions that isolate style changes without being confounded by speaker changes.
To address this issue, we employ Chatterbox \cite{chatterboxtts2025}, a voice conversion model, to adjust the timbre of each generated sample so that it matches the anchor speech while retaining the intended style.

\subsection{Quality Evaluation}
\label{sec:QEval}
To ensure the quality and controllability of the generated speech, we propose a quality assessment process that evaluates generated samples along three aspects: content preservation, style similarity, and speaker identity consistency.
For content intelligibility and accuracy, we employ a widely-used speech recognition model, Whisper-large-v3 \cite{radford2023robust}, to transcribe the generated speech into text. 
This transcription is then compared against the source text transcript to calculate word error rate (WER) for content accuracy and intelligibility verification.
For style alignment, we utilize a Speech Emotion Recognition (SER) model, emotion2vec \cite{ma2023emotion2vec}, to assess style similarity $\textrm{SIM}_\textrm{sty}$ between the generated speech and its style reference speeches.
For speaker consistency, a speaker verification (SV) model ECAPA-TDNN \cite{desplanques2020ecapa} is used to extract speaker embeddings from both the generated and anchor speech samples. 
The cosine similarity $\textrm{SIM}_\textrm{spk}$ between these embeddings is computed to quantitatively assess whether the speaker’s identity is accurately preserved.
Generated samples are retained only when they satisfy all these criteria: WER $<$ 10, $\textrm{SIM}_\textrm{sty}$ $>$ 0.5 , and $\textrm{SIM}_\textrm{spk}$ $>$ 0.5. 

\subsection{Instruction Design}
Through Stages 1 and 2, we obtain high-quality source-target speech pairs in which both source and target speech samples share the same linguistic content and speaker identity but differ in style descriptions. 
The goal of Stage 3 is to generate natural language editing instructions that accurately describe the transformation from the source style to the target style.
Leveraging the language understanding capabilities of Qwen3-8B \cite{bai2023qwen}, we transform the style description into high-quality editing instructions for each source-target speech pair. 
Specifically, we first prompt Qwen3-8B to compare the style descriptions of the source and target speech samples and explicitly identify the differences between them. 
Based on this comparison, we instruct the LLM to formulate a natural language editing instruction using the template: ``Convert the source speech to \texttt{\{target style\}}."
To enhance instruction diversity and avoid repetitive phrasing, we further prompt the model to generate multiple alternative formulations for each transformation, such as ``Convert the source speech to a whispering style with a high-pitched voice." or ``Convert the source speech into a high-pitched whisper."

\section{ISSE Dataset}
\label{sec:pagestyle}

\noindent\textbf{Overview of the ISSE.} 
Using the proposed pipeline, we construct the Instructed Speech Style Editing Dataset (ISSE).
It is an open-source paired speech style dataset comprising approximately 382 hours of speech.
Within it, the ``Real" portion (recorded human speech) amounts to around 90 hours, and the generated speech occupies about 292 hours.
Fig.~\ref{fig:style_Distributions} illustrates the data distribution in ISSE. 
Fig.~\ref{fig:style_Distributions}~(a) shows the distribution of audio length, where most speech samples fall into short-to-medium ranges, providing balanced coverage across durations.
Fig.~\ref{fig:style_Distributions}~(b) illustrates the style distribution, demonstrating that ISSE encompasses a broad variety of speaking styles with sufficient samples per category, in contrast to prior datasets that are limited to a few emotional labels \cite{nguyen2023expresso,zhou2022emotional}.

\noindent\textbf{Quality of the ISSE.} 
Fig.~\ref{fig:style_Distributions} (c)-(e) present the quality evaluation of the ISSE dataset. Specifically, Fig.~\ref{fig:style_Distributions}~(c) reports the quality of intelligibility, measured using ASR-based word error rate to ensure that the generated speeches remain linguistically accurate. Fig.~\ref{fig:style_Distributions}~(d) evaluates the quality of identity preservation through speaker similarity, confirming that speaker characteristics are maintained across paired samples. Fig.~\ref{fig:style_Distributions}~(e) assesses the quality of style preservation, where style similarity against reference speeches demonstrates that stylistic variations are effectively retained. Together, these results highlight that ISSE achieves a balance between intelligibility, identity consistency, and style fidelity.

\section{Benchmark}
\label{sec:typestyle}

In this section, we establish a benchmark for the instruction-guided speech style editing task.
First, we introduce a baseline model, LlasaEdit, built upon \cite{ye2025llasa}.
We then train LlasaEdit separately on ISSE and ESD to demonstrate the effectiveness of the model and the advantages of our proposed dataset in editing style while preserving transcript and speaker identity.
Finally, we validate the generalization of ISSE by evaluating through TTS and expressive speech synthesis (ESS) conditioned on style descriptions.

\subsection{LlasaEdit}
\begin{figure}[t]
    \centering
    \includegraphics[width=\linewidth]{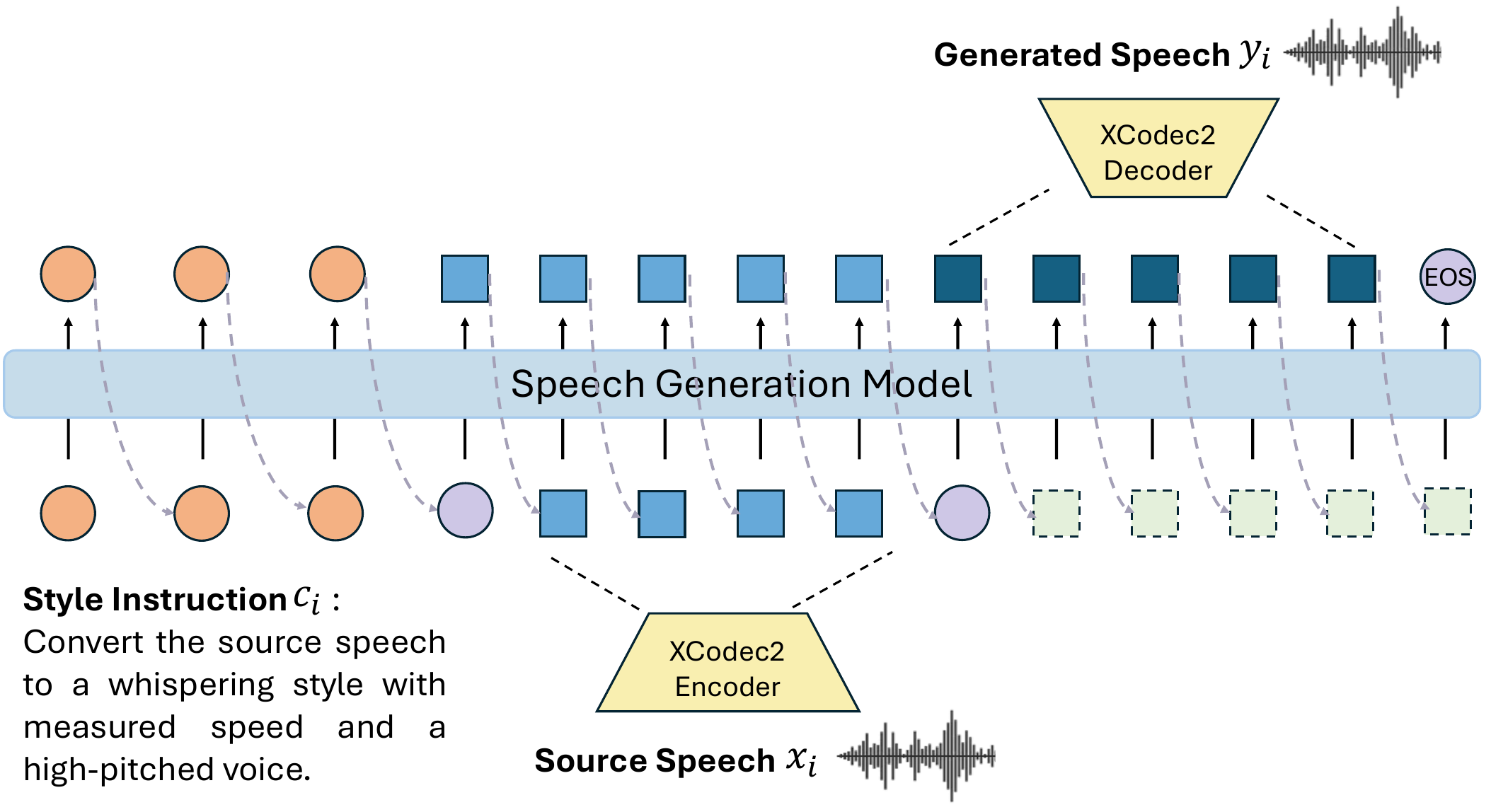}
    \vspace{-10pt}
    \caption{\textbf{The architecture of our baseline LlasaEdit.}}
    \vspace{-10pt}
    \label{fig:speechLLM}
\end{figure}

As shown in Fig.~\ref{fig:speechLLM}, we propose LlasaEdit, which consists of the XCodec2 audio codec and a large transformer-based speech generation model initialized from Llasa~\cite{ye2025llasa}.
Each training instance is represented as a triplet $(x_i, y_i, c_i)$, where $x_i$ denotes the source speech, $y_i$ the target speech (same speaker and transcript but different style), and $c_i$ the corresponding style instruction.
We first encode $x_i$ and $y_i$ into discrete token sequences using XCodec2, while $c_i$ is tokenized into text tokens.
The generation model is then trained to model the conditional distribution of target tokens given the source tokens and the instruction, predicting the target sequence token by token under teacher forcing.
The training objective is to minimize the negative log-likelihood of the target tokens.
In our experiments, we fine-tune the Llasa model~\cite{ye2025llasa} with Low-Rank Adaptation (LoRA)~\cite{hu2022lora}, using rank $r=64$ and $\alpha=128$, applied to both the attention and MLP layers.
We train with AdamW~\cite{loshchilov2018decoupled} (learning rate $1\times 10^{-4}$) for 5 epochs with a batch size of 2, employing learning-rate decay and periodic checkpointing, on 8× NVIDIA RTX A100 GPUs.
At inference, the model takes the source tokens and style instruction, autoregressively generates the edited target tokens, and reconstructs the waveform using the XCodec2 decoder.
For evaluation, we adopt the same metrics as in Quality Evaluation (Section~\ref{sec:QEval}), along with UTMOS~\cite{baba2024utmosv2} for naturalness assessment.

\subsection{Speech Style Editing}
To evaluate the effectiveness of ISSE, we train the LlasaEdit model on the ISSE dataset and compare its performance with the same model trained on ESD, the closest open-source paired speech dataset. Following \cite{huang2024instructspeech}, we convert the categorical labels in ESD into natural language instructions and adopt the same training parameters as used for ISSE. As shown in Table~\ref{tab:tab3}, in the in-domain setting, the model trained and tested on ISSE consistently outperforms its ESD counterpart across all metrics, achieving lower WER (8.06 vs. 10.07), higher $\textrm{SIM}_\textrm{sty}$ (0.68 vs. 0.64), higher $\textrm{SIM}_\textrm{spk}$ (0.58 vs. 0.49), and higher UTMOS (4.29 vs. 4.01). These results highlight the benefits of larger and more diverse data for enhancing generative capabilities. Furthermore, under the cross-domain setting, the model trained on ISSE even surpasses the in-domain performance of the ESD-trained model, with only a marginal shortfall in style similarity. In contrast, the model trained on ESD suffers from severe degradation when evaluated on the ISSE test set, yielding a catastrophic WER of 68.17. 
Those results demonstrate the comprehensive advantages of ISSE over ESD and confirm the effectiveness of its scale, data diversity, and fine-grained style instructions.

\begin{table}[t]
    
    \setlength{\tabcolsep}{4pt}
    \centering
    \caption{\textbf{Results of LlasaEdit trained on different datasets}. LlasaEdit-ESD and LlasaEdit-ISSE denote models trained on ESD and ISSE, respectively.}
    \vspace{5pt}
    \begin{tabular}{cc|cccc}
    \toprule
     Model &Evalset &WER &  $\textrm{SIM}_\textrm{sty}$   & $\textrm{SIM}_\textrm{spk}$ &UTMOS \\
    \midrule
    \multirow{2}{*}{LlasaEdit-ESD} 
        & ESD         & 10.07  &  0.64  &0.49  & 4.01   \\ 
        & ISSE & 68.17  & 0.54  &0.49 & 4.12     \\
    \midrule
    \multirow{2}{*}{LlasaEdit-ISSE} 
        & ESD        & \textbf{7.69}    & 0.59 & 0.51  & \textbf{4.31 }     \\ 
        & ISSE & 8.06   &  \textbf{0.68} & \textbf{0.58}   & 4.29  \\
    \bottomrule
    \end{tabular}
    \vspace{-10pt}
    \label{tab:tab3}
\end{table}

\begin{table}[t]
    \setlength{\tabcolsep}{3pt}
    \centering
    \caption{\textbf{Scalability of ISSE on TTS tasks.} “Codec Rec” indicates speech reconstruction with an audio codec, while “Finetuning” denotes fine-tuning the Llasa-1B model on ISSE.}
    \vspace{5pt}
    \begin{tabular}{cccccc}
    \toprule
      Task & Finetuning & WER & $\textrm{SIM}_\textrm{sty}$  & Gender & UTMOS \\
    \midrule
      Codec Rec (GT) & \wmark &5.22  & -  & 0.97 & 3.83 \\
      \midrule
      \multirow{2}{*}{TTS}  & \wmark & 5.48  & 0.48  & 0.54  &  3.93 \\
        & \ymark  & \textbf{3.74}  &  0.49    & 0.54  &  \textbf{4.03} \\
        \midrule
      ESS  & \ymark  & 5.25  & \textbf{0.60}     & \textbf{0.96}  &  3.94 \\
    \bottomrule
    \end{tabular}
    \vspace{-5pt}
    \label{tab:tab4}
\end{table}

\subsection{Speech Synthesis}
The proposed ISSE can also be extended to TTS tasks by leveraging speech samples paired with their style transcriptions.
As shown in Table~\ref{tab:tab4}, fine-tuning Llasa on the TTS task reduces WER from 5.48 to 3.74 and yields a slight improvement in naturalness.
Moreover, the ESS model conditioned on style descriptions further enhances style controllability, achieving a higher style similarity score (0.60) and gender accuracy (0.96), while maintaining comparable naturalness.

\section{Conclusion}
In this paper, we introduced ISSE, a dataset for instruction-guided speech style editing, constructed through an instructed speech data pipeline with LLM, TTS, and VC technologies.
Benchmarking with our LlasaEdit model demonstrates that ISSE enables accurate and controllable style editing, providing a valuable resource for the speech community. 
Nevertheless, the dataset is limited to English, which may constrain its applicability in multilingual scenarios. 
Future work could extend ISSE to broader languages and richer style dimensions, and explore integration with larger generative models to further advance natural language-guided speech editing.

\bibliographystyle{IEEEbib}
\bibliography{strings,refs}

\end{document}